# Rational Compact Modeling of Transient Forced Laminar Convection

M. N. Sabry, and A. E. Hussin*,
Faculty of Engineering, Ain Shams University, Cairo, Egypt
Mechanical Power Engineering

* Corresponding author:
A. E. Hussin
Email: ahmed_eldeinhussin@eng.asu.edu.eg

## 1 Abstract

Although transient convection is ubiquitous in natural and manmade phenomena, few research works attempted to make a *compact model* for it, altogether, others attempted a compact model that contradicts problem physics. The correct modelling pattern is deduced here analytically for a simple geometry, but it can readily be used for many common applications such as the transient heating of an evacuated solar tube due to temporary cloud shading or for a more precise model of transient building wall heating in a zero-energy building approach. As opposed to *detailed model*, which is based on the governing PDE with a fixed and rigid boundary and initial conditions, usually solved numerically, aiming at obtaining temperatures everywhere and at any time for a prescribed boundary and initial conditions, the *compact model* on the contrary is based on a few simple equations, and aims at giving directly the relation between heat flux and the temperature difference producing it, for any boundary and initial conditions. For transient convection prevails as of today an unphysical approach consisting of modeling it by using a time variant thermal resistance. In this work, starting from the energy PDE, applied to a simple but typical transient forced convection problem, we will get **analytically** the **correct modelling pattern**. This has two main advantages:

- It replaces the classical unphysical approach, with a pair of time constants, one for each important temperature (the fluid bulk and that near the wall), which has an evident physical meaning and can be readily extended to more complex geometries.
- In contrast with the time varying resistance model, one and the same model as deduced here can be reused for any arbitrary time varying function of the input heat flux.

Model validation was made by comparing with a fully blown CFD simulation.

## 2 Introduction

Transient heat transfer by convection is very common, including for instance solar collectors subject to transient incoming flux over the day, or due to temporal cloud shading, or cooling of electronic systems during shift of operating points, or batteries or fuel cells of EVs.. Unfortunately, few research works were devoted to its rational modeling. Although this work considers a simple geometry in order to allow an analytical solution to obtain a rational model, resulting model structure can readily be extended to many common practical problems, including vacuum tube solar collectors and transient heating of building walls for a rational design of a zero-energy building design, etc. There are basically two mathematical modeling strategies. The first is based on governing PDEs (mass, momentum and energy balances), called a '*detailed*' model. Its solution, either analytically or numerically, gives us temperatures, velocities, pressures, and heat fluxes everywhere in the domain, and at any instant of time. At this respect it is regarded as a model having infinite degrees of freedom (DOF). The second has a finite and rather very small DOF, which is called a '*Compact Thermal Model*' (CTM) or a Reduced Order Model (ROM), which is justified by the following fact. Although the detailed model is more precise than CTM, detailed model generates too much data (usually through complicated methods) than is necessary for designers and engineers. The latter users prefer the simple set of equations (sometimes only 1 equation) giving the effect of the most important parameters only on heat transfer rate, at least in initial design phases. Hence, CTMs, although they are less precise, are very popular. However, acknowledging that "everything should be made as simple as possible, but not simpler"[1], some CTMs are too simple to be true, if they contradict problem physics, which may lead to serious errors in some applications. The objective of this work is to avoid this by **obtaining the CTM through an analytical** treatment of the original governing equations. For instance, the so-called 'Heat transfer Coefficient' (HTC) as well as any single thermal resistance in conduction problems, they both assume a local and instantaneous relation between temperature difference at a point and heat flux at that point, which is in flagrant contradiction with the distributive nature of heat transfer due to diffusion. Deficiencies of the single resistor model, in steady state problems, especially for nonuniform heating, have since long been highlighted, including suggesting a better CTM to avoid such problems [2], [4] for steady conduction as well as for steady convection problems [5-7]. Transient heat transfer problems, to which this work is devoted, are significantly less studied in literature. Developed dynamic CTM (DCTMs) are concentrated on conduction problems [8-12] only, while very few addressed DCTM for convection (see below).

Some have concentrated on obtaining the detailed solution analytically [13-16] using different techniques, different approximations, and different series expansions. Some others studied the detailed model either numerically or experimentally [19]. In both cases, it seems that the main objective was the detailed model solution, the compact version was only lightly and wrongly addressed. Because the result was cast in the form of a time dependent thermal resistance or a dimensionless heat flux per unit temperature difference (sort of a Nusselt number) as a function of both space and time. This contradicts problem physics. In fact, transient heat flowing through the modeled object is composed of both a static heat (through a thermal resistance $R_{th}$, its inverse being the Nusselt number) **in addition to** a 'dynamic' heat to 'feed' the thermal capacitance $C_{th}$. While the former heat can be modeled as a function of temperature difference, the latter heat is function of the rate of increase of system stored energy, i.e. the time rate of change of temperature. Hence, a physical sound model should be composed of both thermal resistance(s) $R_{th}$ and thermal capacitance(s) $C_{th}$. A serious disadvantage of the time dependent thermal resistance model is that it is tied to a particular time variation of the heat source. The DCTM developed here can, in contrast, well model **any time variation** of heat sources. Remains to mention a behavior that was noticed in many papers, in particular [23], [24] which is the observation of mainly two time zones, indicating at least 2 time constants, and hence two $C_{th}$. This is natural, since we have two important state variables (bulk temperature and that near the wall), which have different dynamics. It should be noticed that the last two papers have suggested using thermal resistances and capacitances, but the framework was coarse-mesh finite element approach and not a real DCTM. Also, worth mentioning is a rather recent article about transient modeling of heat transfer in a multicomponent system [25]. Conductive elements are modeled by thermal resistances and capacitances but unfortunately, convective components are viewed as an external resistance.

To the authors' knowledge, while many correlations exist for the resistive part, i.e. Nusselt number which prevails at steady state, no such correlation exists for the capacitive part (as will be developed here), which appears only during transient operation. In fact, we cannot just take the modeled object's total mass and multiply it by its heat capacity to get an estimate of the equivalent $C_{th}$ because temperature is never uniform. To keep new ideas about how to treat transient effects clear, they will be presented for a simple but typical case having a simple space dependence, which is that of fully developed flow between parallel plates with transient but uniform flux over the plates. Proposed approach can be generalized to other cases later, after grasping problem physics. Unlike the pioneering work of Graetz [26], which concentrated on the distance required to reach the "spatial" no change zone in a steady state problem, i.e., the fully developed zone, we will not consider the spatial entrance zone to concentrate on the time required to reach the "temporal" steady state, in the fully developed zone.

The paper structure is as follows. Problem description is presented in section 3, followed by, in section 4, a description of how to treat **any** time dependence out of the solution that will be obtained for a time step heating. For better readability, all mathematical details are moved to appendices. In particular, the detailed analytical problem solution is given in 10 Appendix A (a quickly convergent series to facilitate next step), followed by the elaboration of a rational DCTM out of it in section 5, together with a rapid discussion of the physical insight it allows. The resulting DCTM will be validated by comparing its predictions with the results of a fully blown CFD model in section 6. Finally, conclusions will be drawn in section 6.3.

---

[1] An edited and paraphrased sentence of Einstein in J*ournal of the Franklin Institute.* 1936.

## 3 Problem description
### 3.1 General
Problem geometry and boundary conditions are shown in Figure 1.

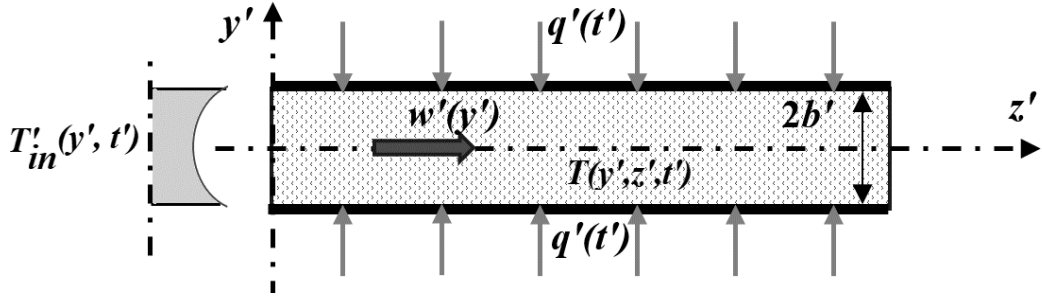

*Figure 1- Problem geometry and boundary conditions*

In the sequel, all dimensional quantities will have the ' sign. Fluid domain is delimited by two parallel planes normal to the $y'$-axis, that are apart by a distance $2b'$. Fluid flows between the plates in a direction parallel to the $z'$-axis in the laminar regime. Plate width in the x'-direction (normal to the figure) is $d'$. There is complete symmetry in the $x'$-axis, i.e., the problem is 2D. Fluid velocity is assumed to be fully developed and steady $w'(y')$, but not temperature $T'$, which is assumed transient $T'(y', z', t')$, where $t'$ stands for time. Fluid receives a heat flux through walls that is uniform in space but is time dependent $q'(t')$. Fluid inlet temperature is given $T'_{in}(y', t')$. The initial condition is also given and takes the following form in general $T'_0(y', z')$.

### 3.2 Dimensional analysis
The following characteristic dimensions are assumed to simplify resulting equations. Characteristic length $L_{ch}$ is duct half height $b'$, and characteristic velocity $V_{ch}$ is the average flow velocity. Characteristic heat flux $q_{ch}$ is derived from the value of the imposed uniform heat flux $q'$, while characteristic temperature difference $DT_{ch}$ is defined such as to satisfy: $DT_{ch} = q_{ch} L_{ch} / k'$ (where $k'$ is the thermal conductivity). Finally characteristic time $t_{ch}$ is defined such as to satisfy $t_{ch}=L_{ch}^2/a'$ (where $a'$ is the thermal diffusivity). Nondimensional velocity $w$, coordinates $y$ and $z$ and time $t$ will all have the same symbol as their dimensional counterpart without the dash, except nondimensional temperature, which is $q = (T'' - T'_{ref})/DT_{ch}$, where $T'_{ref}$ is an arbitrary reference.

### 3.3 Hydrodynamic problem
Since the hydrodynamic part of the problem is steady and fully developed, there is only one nondimensional velocity component $w$ in the $z$ direction, which is function of nondimensional $y$:

$$w(y) = \left(\frac{3}{2}\right)(1 - y^2) \tag{1}$$

In case a flat velocity profile was assumed, $w(y) = 1$

### 3.4 Thermal problem
The governing equation is the energy equation, in its nondimensional form:

$$\frac{\partial \theta}{\partial t} + \text{Pe } w \frac{\partial \theta}{\partial z} = \frac{\partial^2 \theta}{\partial y^2} + \frac{\partial^2 \theta}{\partial z^2} \tag{2}$$

where Pe is the Peclet number Pe = $V_{ch} L_{ch}/a'$. Space dependence of boundary conditions both at the wall and at inlet will be assumed as simple as possible to concentrate on the main issue here, which is the time dependence. Hence the wall boundary condition is:

$$\left.\frac{\partial \theta}{\partial y}\right|_{y=\pm 1} = q(t) \tag{3}$$

where $q(t)$ is any arbitrarily given nondimensional time function.

Inlet boundary condition deserves some scrutiny. While we can theoretically impose any profile, it should not be in contradiction with other conditions. Suppose we impose a flat profile for instance. At entry section near the wall, the point ($z=0$, $y \rightarrow 1$), we will have a contradiction between temperature being flat, hence $\partial q/\partial y=0$ and the boundary condition (3). This would create an undesirable discontinuity that will be avoided here as explained below.

For a steady state problem, we have the right to impose any "physically realizable" inlet temperature profile respecting constraints of the previous paragraph. In general, it will create a thermally developing zone before reaching the thermally fully developed regime. This problem has been studied by many authors since the pioneering work of Graetz [26]. Most of these studies considered uniform wall boundary conditions. This has been recently generalized to nonuniform wall boundary conditions [5-7], after generalizing the concept of thermally fully developed. In both cases, uniform or nonuniform wall boundary conditions, calculating temperature field in the developing zone is quite involved. Adding transient effects in this zone would needlessly complicate the problem. Our purpose is to understand some basic physical features related to dynamic effects, not to get lost in endless mathematics. To avoid discontinuity mentioned in previous paragraph and further simplify the problem, the role of inlet conditions will be neutralized in a first approach. Inlet temperature profile **is assumed to be the same as fully developed steady state one** to avoid the thermally developing zone. Of course, this can be relaxed later after reaching an understanding of the physics of dynamic effects.

Initial boundary condition will be simply:

$$\theta(y, z, t = 0) = \theta_0(y, z) \tag{4}$$

where $q_0(y, z)$ is any arbitrarily given space function.

## 4 Time variation of surface heat flux

It is required to get a model for the dynamic response of a system due to any given time variation of $q(t)$. The simplest form of time variation of the imposed surface heat flux is:

$$q(t) = q(t)|_{US} = u(t) \tag{5}$$

where $u(t)$ is the Heaviside unit step function, and subscript *US* stands for unit step. For better readability, the corresponding dynamic temperature field $q(y, z, t)|_{US}$ will be obtained in 10 Appendix A 10Appendix A. Starting from it we can easily get the impulse response $\partial q_{US}/\partial t$, which is the temperature field due to a unit impulse in time, and hence the response due to any other arbitrary $q(t)$ is:

$$\theta(y, z, t) = \int_{\tau=0}^{t} \frac{\partial \theta(y, z, \tau)|_{US}}{\partial \tau} q(t - \tau) d\tau \tag{6}$$

Use will also be made of the Laplace transform to express this general model in the form of thermal bipolar elements (resistances and capacitances) for an easy interpretation of the physical meaning of the obtained general model.

## 5 Dynamic Compact Thermal Model
### 5.1 Building the Compact Model

Detailed temperature field was obtained in10 Appendix A for a uniform wall heat flux in space and a unit step in time. It gives us the Laplace transform $\overline{\theta}(y, z, s)\big|_{US}$ due to $\overline{q}(s)|_{US} = \frac{1}{s}$ where the subscript *US* stands for a unit step heat input in time. The Laplace variable is $s$, while the overbar over a function denotes the Laplace transform of that function. Due to linearity, the temperature field due to any other time variation of wall heat flux, provided it remains uniform in space, having the Laplace transform $\overline{q}(s)$ is simply (which is also the Laplace transform of (6):

$$\overline{\theta}(y, z, s) = \left(s\overline{\theta}(y, z, s)\big|_{US}\right) \overline{q}(s) \tag{7}$$

Let us exploit this interesting and quite general result further. Engineers designing heat transfer equipment are only interested by the following transfer functions. First, the transient bulk temperature $q_b$ (by simply averaging (A-17) over $y$, taking $w$ as the weight) satisfies:

$$\frac{\theta_b(z,s)}{\overline{q}(s)} = \theta_{in,b} + \left(\frac{1}{s} - \theta_{in,b}\right)\beta_0(z,s) \tag{8}$$

$$\beta_n(z,s) \triangleq 1 - e^{-m_n z} \tag{9}$$

And second the thermal impedance $Z_{th}$, i.e., the ratio of (wall – bulk temperature) to the heat flux:

$$Z_{th}(z,s) \triangleq \frac{\overline{\theta_w} - \overline{\theta_b}}{\overline{q}} = R_{ss} - \sum_{n=1}^{\infty}\left[\frac{2s\beta_n(z,s)}{n^2\pi^2(n^2\pi^2 + s)}\right] \tag{10}$$

Where $R_{ss} = 1/3$. In order to proceed further, we need to approximate $b_n$ to elucidate major physical aspects blurred by the complicated expression of $m_n$. Let us approximate the exponent $m_n z$ and use Pade approximation of order 1 for the exponential to get:

$$\beta_n(z,s) \simeq \frac{\left[\frac{(n^2\pi^2 + s)}{Pe^2}\right](Pe\,z)}{1 + \left[\frac{(n^2\pi^2 + s)}{Pe^2}\right](Pe\,z)} \tag{11}$$

The error between the exact (9) and approximate form (11) of $b_n$ is plotted in Figure 2. For large values of $s$ (small values of time) the error is very small (less than 5%, tending to 0 as $s\to\infty$), but for smaller values of $s$ (larger values of time) the error increases before falling out to zero at steady state ($s\to 0$), but it remains rather bounded for intermediate times. In fact, even at intermediate values of $s$, the error is very small at both $z\to 0$, $z\to\infty$. This is a remarkable approximation. It will directly lead us to a simple and rather accurate dynamic compact thermal model.

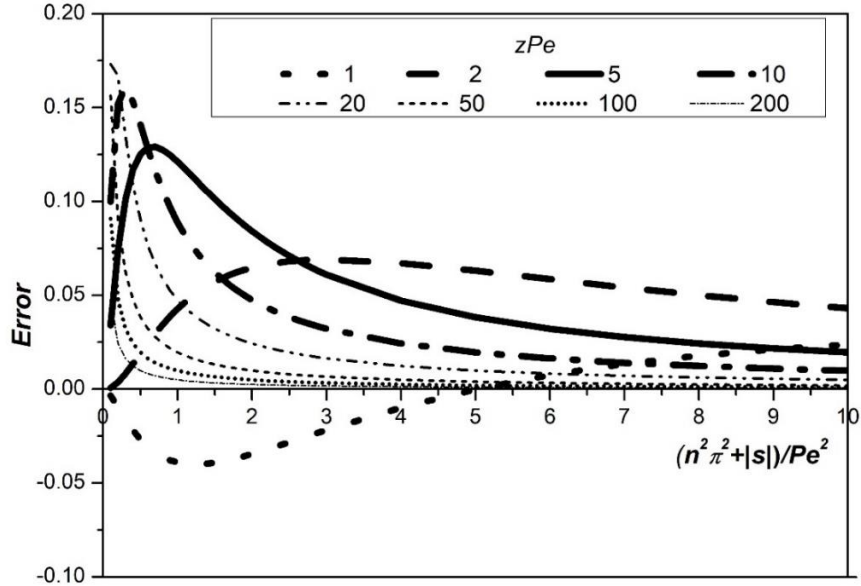

Figure 2- Error in approximating $b_n$

Using the approximation (11) and taking only one term in (10), both transfer functions are

$$\frac{\theta_b(z,s)}{\overline{q}(s)} = \theta_{in,b} + \left(\frac{1}{s} - \theta_{in,b}\right)\frac{s\frac{z}{Pe}}{1 + s\frac{z}{Pe}} = \frac{\theta_{in,b} + \frac{z}{Pe}}{1 + s\frac{z}{Pe}} \tag{12}$$

$$Z_{th}(z,s) = R_{ss} - \frac{2s\frac{z}{Pe}}{\pi^2\left(1 + (\pi^2 + s)\frac{z}{Pe}\right)} = \left(R_{ss} - \frac{2}{\pi^2}\right) + \frac{\frac{2}{\pi^2}}{1 + s\frac{\frac{z}{Pe}}{1+\pi^2\frac{z}{Pe}}} \tag{13}$$

They can now be schematized by an equivalent thermal network Figure 3.

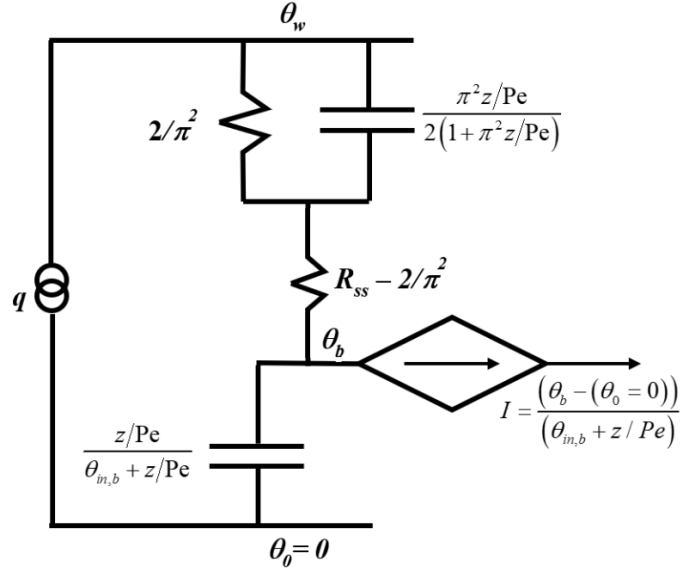

Figure 3 – Full thermal equivalent circuit in lumped dimensionless form

Please note that the equivalent circuit shown in Figure 3 is the result of many approximations that were analytically shown to have little effect on predicted physical behavior 10 Appendix B). Flat velocity profile was assumed, only one term in the series (10) was retained and the approximation for $b_n$ (11) was assumed to hold. Also, inlet temperature profile was assumed to be that of steady state fully developed zone. The resulting equivalent circuit obtained in Figure 3 clearly explains physical behavior, at least qualitatively, as will be elaborated in the next subsection. It is quite possible to avoid the above-mentioned approximations, but quite messy algebra would result, as a consequence a needlessly more complicated equivalent circuit would result. It would modify the result a bit, while keeping it at least qualitatively similar. The clear physical interpretation of this simple DCTM is a plus. The obtained DCTM will be validated by comparing its predictions with those of a fully blown CFD model (section 6).

5.2   Physical interpretation of the compact model built

Thermal capacitances represent energy storage in the system during initial heating (if the system was heated) to reach a steady state, if any (i.e. if the rate of heating stabilizes with time). Obviously, we need at least two capacitances, which is the case in this DCTM, to track the dynamics of both bulk temperature and wall temperature. The temperature (equivalent to voltage) controlled heat (equivalent to current) source (or sink) represents energy (enthalpy) getting away from the system with fluid outflow. All circuit elements (heat sources, thermal resistances and capacitances) are in fact distributed elements per unit area of the heating surface. Circuit elements are expressed in dimensionless form. They are represented in a lumped form here for ease of understanding.

Note that at steady state (subscript $ss$) (corresponding to $s \to 0$, i.e., by removing capacitances) this simple DCTM will yield the well-known steady state results for both bulk and wall temperatures. In particular, at steady state (if $q'(t')$ stabilizes), at the bulk temperature node incoming heat exactly counterbalances outgoing heat through the outgoing temperature-controlled heat source (outgoing enthalpy) i.e. the First Law is satisfied, which in a dimensional form reads:

$$T'_{ss}(z) - T'_{ref} = (T'_{in,b} - T'_{ref}) + z' \frac{q'_{ss}}{\rho c_p b' V_{char}} \qquad (14)$$

As for the difference between wall and bulk temperatures, it depends on the steady state thermal resistance $R_{SS}$, which is the inverse of the steady state heat transfer coefficient HTC, in dimensionless form. The latter is the well-known Nusselt number, which is 12 if the characteristic length was the hydraulic diameter = $4b'$. For the characteristic length chosen here, channel half depth, i.e. $b'$, Nusselt number is 3, and hence, as expected, $R_{ss}=1/3$.

Obtained DCTM provides, in addition, full information about the dynamics of both bulk and wall temperatures at any $z$ before reaching steady state through both capacitances. Each capacitance is a distributed one along the whole duct, going from 0 at inlet to a 'fully developed' value at large values of $z$. The 0 initial value is a consequence of the boundary condition imposed, which is a steady state inlet temperature profile, i.e. dynamics at inlet were intentionally suppressed to let analytical solution be easy to obtain. Another boundary condition could have also been imposed, giving rise to a more complicated equivalent circuit, which has been deliberately avoided here. The 'fully developed' value of thermal capacitances is 1 for the bulk and 0.5 for the wall – bulk, both dimensionless, knowing that the unit dimension of the capacitance per unit area of the heating surface is $r\,c_p\,b'$. Note that the crude approach of taking as a capacitance a lumped value of the whole fluid mass multiplied by its thermal capacitance, gives only a partial and quite approximate view. The current analysis elucidated the bigger picture and gave more exact results in multiple respects. First, we have two capacitances instead of one. Second, it is a distributed value not a lumped one. Finally, capacitance value is function of $z$, in our boundary conditions it gradually increases from 0 to the 'fully developed value'.

## 6 Comparison with a CFD model

To validate the proposed Dynamic Compact Thermal Model (DCTM), or equivalently the proposed equivalent thermal network, we conducted simulations of transient forced laminar convection between two-parallel-plates. The DCTM results were compared with predictions of Computational Fluid Dynamics (CFD) to validate different aspects detailed below.

Comparisons were made for the following conditions: the plates were 0.2 m long and 0.01 m apart, with a laminar fluid flow (for different fluids) with a flat velocity profile (case of slip condition) of 0.01 m/s, or parabolic (case of no-slip condition) with the same average velocity. The inlet temperature profile satisfies (A-8) for no slip and (A-9) for the slip condition. Unless otherwise stated, heating was a step function at the initial time of 100W/m$^2$ (other time varying heating rates were explored), and fluid is air (other fluids were considered). It is worth mentioning that the fluid properties were kept constant in the CFD model to fulfill the assumptions of DCTM. The steady state bulk and wall temperatures were used as indication of the accuracy of the DCTM model. Different mesh sizes were examined, and it is found after many trials that a structured mech with mesh size of 0.1 mm in $y$ and $z$ directions is adequate. Furthermore, the time-step dependence test for the CFD model shows that a time-step of 0.1 s is quite sufficient.

Results have shown the same trends for all cases treated, which will be summarized in section 6.2. Clearly, proposed DCTM predictions perfectly matched CFD results for both very small and very large times. An error is observed for intermediate times, although it remains bounded. This was predictable (see section 5), due to the 1$^{st}$ order Pade approximation. Second or higher order Pade approximations could have been made, to get a more precise model. This would have resulted in a more complicated thermal network, which goes against the objective of any compact model: very simple and handy, although never 100% accurate.

### 6.1 Obtaining results of the compact model

To obtain temperatures as a function of time out of the equivalent circuit, we only need to solve 2 similar and uncoupled ordinary differential equations (ODEs) each of them is of the first order having the following common form: ($q$ is either bulk temperature or wall - bulk, $R$ is resistance, $t$ is the time constant (i.e., the product $RC$), $t$ is time, $q_m$ is characteristic heat flux, and $r(t)$ is the dimensionless time rate of change of the imposed heat flux, $t$, $R$, $q_m$, $q_0$ are given constants).

$$\tau \frac{d\theta}{dt} + \theta = R\,q_m r(t) \tag{15}$$

$$\theta(0) = \theta_0 \tag{16}$$

The externally applied transient heat flux function (i.e. $r(t)$ in its dimensionless form) must be given to solve the ODE. The model can handle any input function. In the sequel, it will be validated for only two typical functions: a step input or a ramp input for a certain time $T$, followed by a constant value. Both cases can be grouped here in only one as follows:

$$r(t) = \begin{cases} t/T & if\ t \leq T \\ 1 & if\ t > T \end{cases} \tag{17}$$

Eventually, letting $T \to 0$ gives the step input. The general solution is as follows:

- For $t \leq T$

$$\theta(t) = (\theta_0 + R\,q_m\tau/T)e^{-t/\tau} + R\,q_m(t-\tau)/T \tag{18}$$

$$\theta(t=T) = \theta_1 = (\theta_0 + R\,q_m\tau/T)e^{-T/\tau} + R\,q_m(T-\tau)/T \tag{19}$$

- For $t \geq T$

$$\theta(t) = (\theta_1 - R\,q_m)e^{-(t-T)/\tau} + R\,q_m \tag{20}$$

- In case $T \to 0$; first part $t \leq T$ is irrelevant, second part $t \geq T$ becomes:

$$\theta_1 = \theta_0 \tag{21}$$

$$\theta(t) = (\theta_0 - R\,q_m)e^{-t/\tau} + R\,q_m \tag{22}$$

## 6.2 Validation of results of the compact model for different usage scenarios

### 6.2.1 Validation for different velocity profiles

An important approximation was made to obtain the DCTM, which is that of assuming a flat velocity profile, that is why it is interesting to see how the DCTM would behave for both flat and parabolic velocity profiles. Obtained thermal network (the DCTM) contained the thermal resistance $R_{SS}$, which is the dimensionless steady state thermal resistance between wall and bulk temperatures. It is simply the inverse of the Nu number. Since our characteristic length is half the channel depth instead of the commonly used hydraulic diameter (double the channel depth), our $R_{SS}$ is 4 times greater than the commonly used values. For slip velocity profile it is 1/3 while for no-slip case it is 1/2. To be fair, comparison should be made taking in each case the appropriate value of $R_{SS}$ in the DCTM, which gives results shown in Figure 4 and Figure 5 for air and a unit step heat addition, with slip and no-slip conditions respectively.

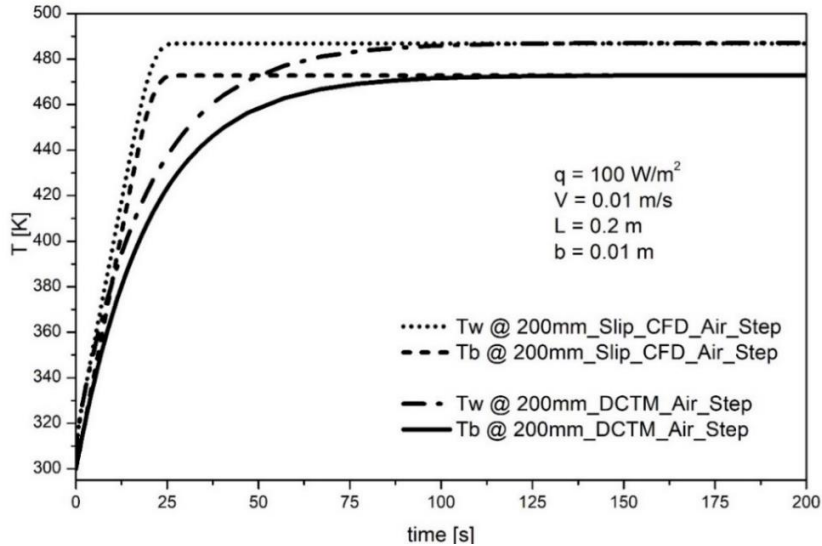

*Figure 4: Comparison of wall and bulk temperature profiles for air between DCTM and CFD model (case of unit-step surface heat flux of 100 W/m², slip condition with 0.01 m/s of air, and thermally fully developed inlet temperature profile).*

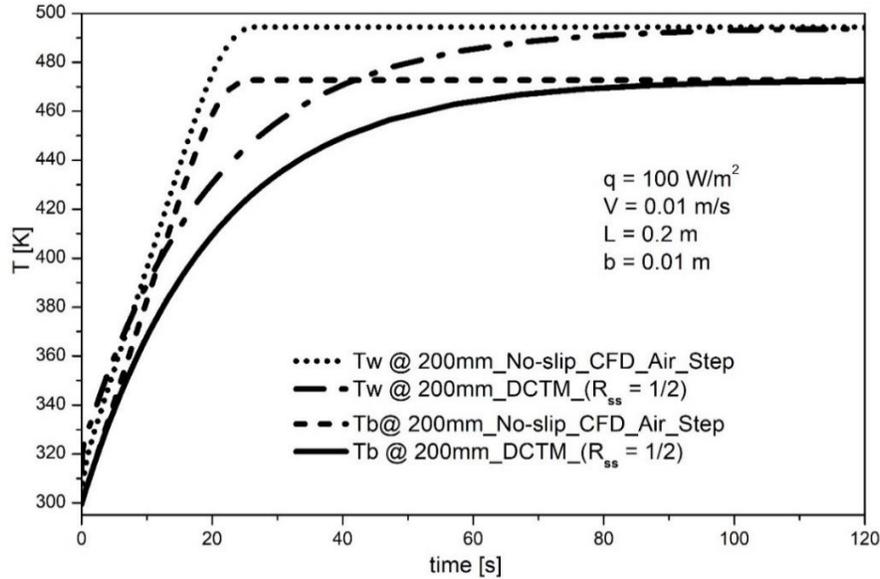

*Figure 5: Comparison of wall and bulk temperature profiles for air between DCTM and CFD model (case of unit-step surface heat flux of 100 W/m², no-slip condition with an average of 0.01 m/s of air, and thermally fully developed inlet temperature profile).*

#### 6.2.2 Validation for different time rates of heating

It is also important to show that the proposed DCTM would predict correct values, regardless of the shape of time rate of heating. For this, heating for the case of a ramp going from 0 to 100 W/m² over 25s, followed by a flat rate, was also tested, giving results shown in Figure 6.

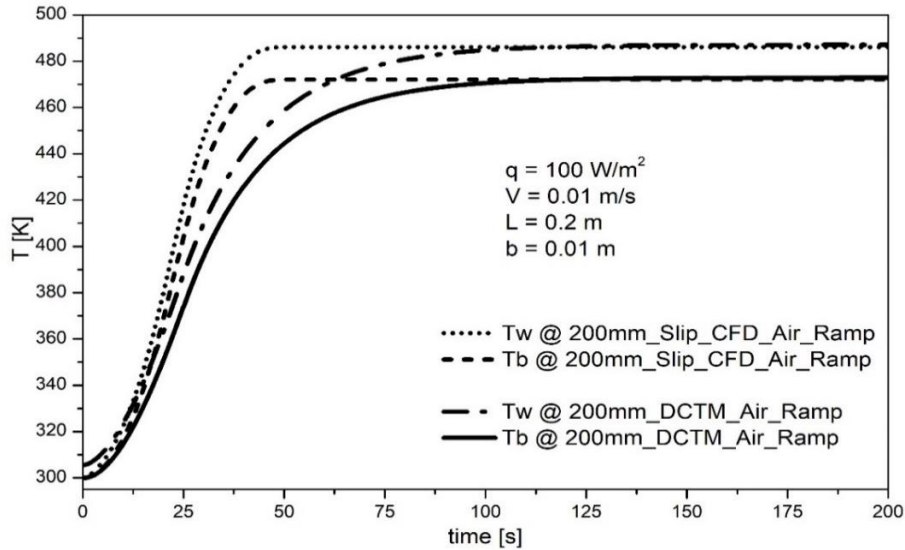

*Figure 6: Comparison of wall and bulk temperature variations between DCTM and CFD model (case of ramp surface heat flux of 100 W/m² over 25s, slip condition with 0.01 m/s of air, and thermally fully developed inlet temperature profile).*

#### 6.2.3 Validation for different fluids

Moreover, the DCTM's performance was evaluated with various working fluids, including hydrogen ($H_2$), ammonia ($NH_3$), and water ($H_2O$). As depicted in Figure 7

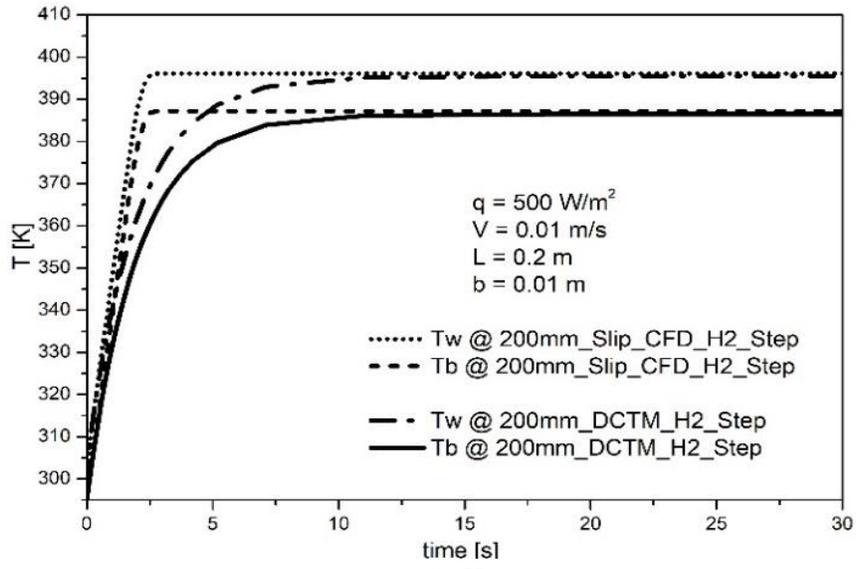

(a) H₂

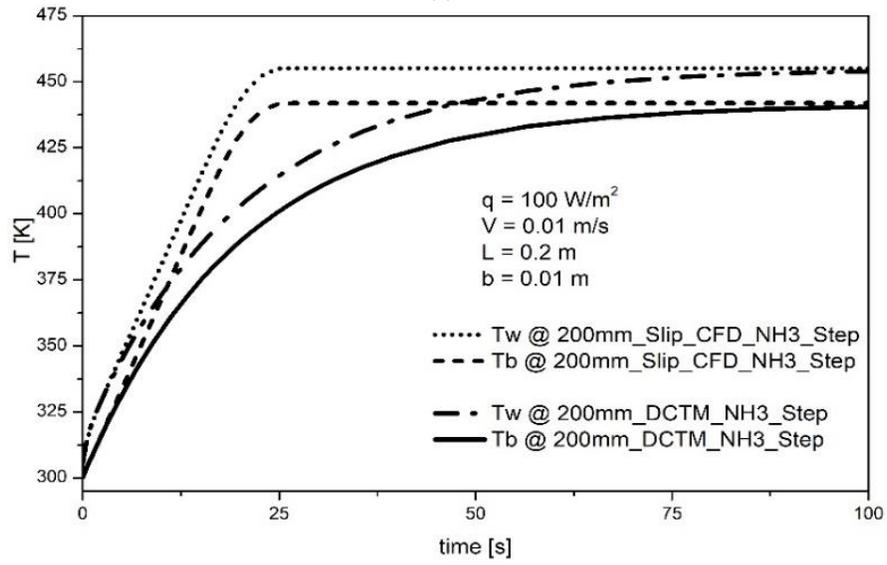

(b) NH₃

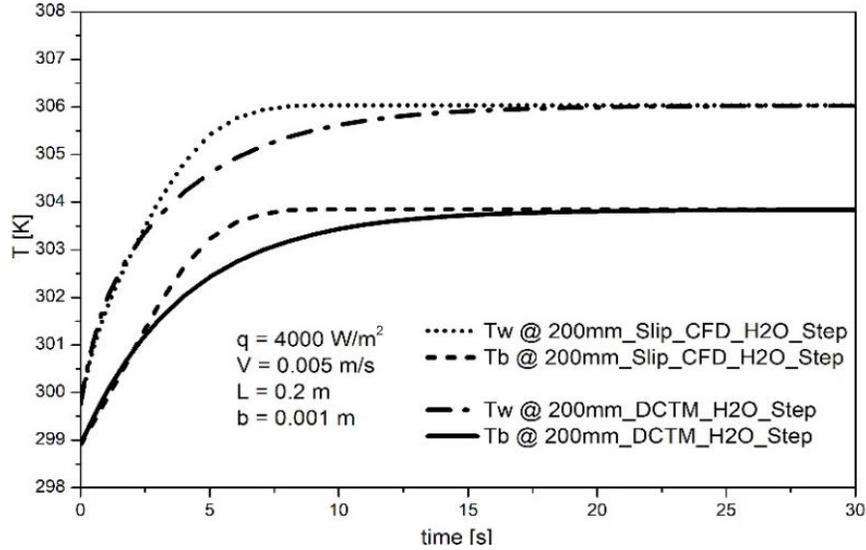
(c) $H_2O$

*Figure 7: Comparison of wall and bulk temperature variations between DCTM and CFD model (case of unit-step surface heat flux, thermally fully developed inlet temperature profile, and slip condition or (a) $H_2$, (b) $NH_3$, and (c) $H_2O$.*

### 6.3 The relative merits of the proposed compact dynamic thermal model DCTM

The main advantage of DCTM compared to a fully blown CFD model, is that DCTM has given us a direct insight into the underlying physics, elucidating the dynamics of both bulk and wall temperatures as opposed to CFD results, which are just a large quantity of data coming without any self-explanations. Apart from that, DCTM is highly more efficient than CFD. The latter requires the work of a specialist, adjusting mesh size in both directions, as well as time step sizes over many iterations consuming many hours or even more before a mesh size independent result is obtained. As for DCTM, it only requires solving 2 similar and uncoupled ODEs of the first order each, and with constant coefficients, which is within the reach of any unexperienced person having a minimum scientific background in a matter of minutes. Even without any scientific background, one can just enter the equivalent circuit in a dedicated SW package (such as MATLAB® for instance) to obtain the result in almost no time. Obviously, CFD is more accurate, which is a common feature of all compact models: fast and handy in initial design phases, requiring many iterations, and as long as accuracy is acceptable, the compact model will always be the tool that is most often used.

### 7 Conclusion

A new and rational modeling approach has been proposed for transient forced convection. Previous approaches were mainly based on a time dependent Nusselt number, which is in contradiction with problem physics. Fluid has a thermal capacitance, which is not captured in the classical approach. This capacitance will consume part of the supplied heat while heating. This heat cannot be modeled as heat flowing through a resistance, i.e. an element where heat flow is proportional to the temperature difference. But rather as heat flowing into a capacitance where heat flow is proportional to the time rate of increase of system stored energy, i.e., the time rate of increase of temperature. Obtained model structure can readily be reused for many practical applications.

The methodology used was based on first solving governing equations of the detailed model for a simple problem to obtain the temperature field everywhere and at any moment. Problem simplicity has allowed an analytical solution, after introducing some simplifying assumptions that were analytically proved to have a small effect. Out of this solution, an extremely simple dynamic compact model was deduced, cast in the form of an equivalent thermal network of a pair of distributed in space but constant in time thermal resistances and capacitances. Values of these elements were deduced as a function of fluid properties, domain size and location of temperature evaluation. Main advantages of the proposed approach over the classical one, based on a time varying thermal resistance with no capacitive effect whatsoever, are:
- It gives the correct pattern yielding two time-constant, one for each important state variable: fluid bulk temperature and that near the wall. Other more complicated problems could be modeled using this same pattern.
- One and the same model obtained using this approach can predict system behavior due to any other time function of the applied source.

To assess the validity of the assumptions made, a CFD model was constructed and solved to compare predictions of the Dynamic Thermal Compact Model (DCTM, in our case, the thermal network of resistances and capacitances) with those of the CFD model. Results were highly satisfactory. As expected by the analytical treatment of the major simplifying assumption made (using 1$^{st}$ order Pade approximation of an exponential function), the error tends to zero both at very small and very large times. For intermediate times, the error remains bounded.

## 8 Funding

The authors declare that no funds, grants, or other support were received during the preparation of this manuscript.

## 10 Appendices

### Appendix A    Solution of the thermal problem for a unit step input

*During this whole section the temperature field q<sub>US</sub>(y, z, t), due to a unit step surface heat flux, will be deduced. Hence, for better readability the subscript US will be dropped.*

#### A-1    Problem splitting

Let us split the transient temperature distribution into two functions, each satisfying part of problem constraints, but the sum of both satisfies all problem constraints:

$$\theta(y,z,t) = \theta_{QS}(y,z,t) + \theta_{IE}(y,z,t) \tag{A-1}$$

where $q_{QS}$, in which the subscript *QS* stands for Quasi Static, satisfies wall and inlet boundary conditions, including their time variation, but does not take into consideration the time rate of change of system stored energy. It satisfies the following equations:

$$\text{Pe } w \frac{\partial \theta_{QS}}{\partial z} = \frac{\partial^2 \theta_{QS}}{\partial y^2} + \frac{\partial^2 \theta_{QS}}{\partial z^2} \tag{A-2}$$

$$\left.\frac{\partial \theta_{QS}}{\partial y}\right|_{y=\pm 1} = q(t) \tag{A-3}$$

Inlet boundary condition will be adjusted to avoid a "developing" zone, as will be explained below. As for the second component, $q_{IE}$, where the subscript *IE* stands for Inertial Effect, its main role is to take into consideration the time rate of change of system stored energy, hence initial conditions as well. Wall and inlet conditions are zero because they were handled by $q_{QS}$. The field $q_{IE}$ satisfies:

$$\frac{\partial \theta_{IE}}{\partial t} + \text{Pe } w \frac{\partial \theta_{IE}}{\partial z} - \frac{\partial^2 \theta_{IE}}{\partial y^2} - \frac{\partial^2 \theta_{IW}}{\partial z^2} = -\frac{\partial \theta_{QS}}{\partial t} \tag{A-4}$$

$$\left.\frac{\partial \theta_{IE}}{\partial y}\right|_{y=\pm 1} = 0 \tag{A-5}$$

$$\theta_{IE}|_{t=0} = \theta_0 - \theta_{QS}|_{t=0} \tag{A-6}$$

$$\theta_{IE}|_{z=0} = 0 \tag{A-7}$$

The time rate of change of system stored energy, initially neglected by the $q_{QS}$ part, has not disappeared, it reappears in the equations for the $q_{IE}$ part as a volumetric heat source. It is also readily seen that the sum of the two fields $q_{QS}+q_{IE}$ satisfies all problem constraints.

#### A-2    Solution for the Quasi Static Part

The solution of the Quasi Static problem can be directly derived from the steady problem, after a minor modification. For a parabolic velocity profile, we have:

$$\theta_{QS}(y,z,t) = u(t)\left[\left(\frac{3y^2}{4} - \frac{y^4}{8} - \frac{39}{280} + \theta_{in,b}\right) + \frac{z}{\text{Pe}}\right] \tag{A-8}$$

where $q_{in,b}$ is the arbitrarily given bulk temperature at inlet. For a flat velocity profile, we have:

$$\theta_{QS}(y,z,t) = u(t)\left[\left(\frac{y^2}{2} - \frac{1}{6} + \theta_{in,b}\right) + \frac{z}{\text{Pe}}\right] \tag{A-9}$$

The expression (A-8) (as well as (A-9) for a flat velocity profile) already satisfies (A-2) and (A-3). In addition, bulk temperature at inlet $q_{in,b}$ is a constant that can be adjusted to match any given inlet condition. The details of the inlet temperature profile $q_{in}(y, t)$ can be arbitrary. If it does not match $q_{QS}(y, 0, t)$, then a thermally developing zone will form at duct inlet, after which the solution (A-8) will be valid. The thermally developing zone was treated elsewhere for steady state [5]. As explained in section 3.4, we wish to avoid it in this work. That is why for simplicity, assume:

$$\theta_{in}(y,t) = \theta_{QS}(y,0,t) \tag{A-10}$$

## A-3    Solution Methodology for the Inertial Effect Part

The required field $q_{IE}(y, z, t)$ should satisfy (A-4) to (A-7), (inlet condition was satisfied by $q_{QS}$ (A-10). It is well known that the initial condition (A-6) can be replaced by a volumetric heating one, provided that the value at the RHS of (A-6) was added to the RHS of (A-4) after multiplying it by the Dirac distribution $d(t)$ (see below (A-11)). The problem will then be that of *a zero-wall condition*, *a zero-inlet condition* and *a zero-initial condition*. It will only be driven by a non-zero "volumetric heating" term, which is the modified RHS of (A-4). To state this fact in a clearer form, let us take the Laplace transform of **(A-4)**:

$$\text{Pe } w \frac{\partial \bar{\theta}_{IE}}{\partial z} - \frac{\partial^2 \bar{\theta}_{IE}}{\partial y^2} - \frac{\partial^2 \bar{\theta}_{IE}}{\partial z^2} + s\bar{\theta}_{IE} = -s\bar{\theta}_{QS} + \left(\theta_0 - \theta_{QS}\big|_{t=0}\right) \tag{A-11}$$

Laplace variable is $s$, Laplace functions are denoted by an overbar. All conditions are homogeneous on the wall ($y=\pm 1$) and at inlet ($z=0$). Hence, we will attempt a solution in the form of a series expansion over the following functions of $y$, which identically satisfy wall boundary conditions (for simplicity, in the sequel the second term in the RHS of (A-11) is assumed 0):

$$\bar{\theta}_{IE}(y, z, s) = \sum_{m=0}^{NT} \cos(m\pi y) f_m(z, s) \tag{A-12}$$

where $NT$ is the retained number of terms in the space series expansion of the temperature. Functions $f_m(z, s)$ are the remaining unknowns. The last boundary condition to be satisfied is to let functions $f_m(z=0, s) = 0$. Substituting (A-12) in (A-11), multiplying by $[2/(1 + d_{n0})] \cos(n p y)$ and integrating over $y$ from 0 to 1, gives the ODE set:

$$D^2 f_n - A_n f_n - \text{Pe} \sum_m B_{mn} D f_m = S_n \tag{A-13}$$

where the symbol $D$ denotes differentiation wrt $z$ and coefficients $A_n$, $B_{mn}$, and $S_n$ are given in Appendix B. Having got $A_n$, $B_{mn}$ and $S_n$, we can readily solve the ODE (A-13) to get $f_n$, and substitute in (A-12) to get the Laplace transform of the solution for the dynamic part. Required boundary conditions for $f_n$ are zero at inlet ($z=0$) and boundedness as $z \to \infty$ (except for the inevitable linear increase term due to uniform heating). Since for the two cases, flat and parabolic velocity profiles, we have shown in Appendix B Appendix B that the numbers $A_n$, $B_{mn}$ and $S_n$ are rather close, we are not expecting a fundamentally different solution. For flat velocity profile, $f_n$ are decoupled for different values of $m$, and hence can easily be obtained. For a parabolic velocity profile, coupling exists although it is rather weak. This will only let algebra required to get $f_n$ become more involved, without adding fundamental new aspects to thermal dynamics. That is why in the sequel, we will only concentrate on the flat velocity profile to better understand problem physics, numerical verification in section 6 will show that this simplification is valid

## A-4    The Temperature Field for a Flat Velocity Profile

The set of equations **(A-13)** are all linear ODEs of the second order with constant coefficients. Moreover, they are decoupled: recall that for a flat velocity profile $B_{mn}$ is the identity matrix. Boundary conditions were stated above. The solution is thus simple by standard methods giving:

$$f_0 = \left(\frac{1}{s^2} - \frac{\theta_{in,b}}{s}\right)(1 - e^{-m_0 z}) - \frac{z}{s\text{Pe}} \tag{A-14}$$

$$f_{n>0} = \frac{2(-1)^{n+1}}{n^2 \pi^2 (n^2 \pi^2 + s)} (1 - e^{-m_n z}) \tag{A-15}$$

where:

$$m_n = \frac{\left(\sqrt{\text{Pe}^2 + 4(n^2 \pi^2 + s)} - \text{Pe}\right)}{2} \tag{A-16}$$

Substituting in **(A-12)** and hence **(A-1)**, using (A-9), gives us the full dimensionless temperature field:

$$\bar{\theta}(y,z,s) = \frac{\left(\frac{y^2}{2} - \frac{1}{6} + \theta_{in,b}\right)}{s} + \left(\frac{1}{s^2} - \frac{\theta_{in,b}}{s}\right)(1 - e^{-m_0 z})$$
$$+ \sum_{n=1}^{\infty} \cos(n\pi y) \left[\frac{2(-1)^{n+1}}{n^2\pi^2(n^2\pi^2 + s)}(1 - e^{-m_n z})\right]$$
(A-17)

It is worthwhile noting that, as expected, taking the limit: $\lim_{s \to 0}(s\bar{\theta})$ gives us the expected steady state solution, while $\lim_{s \to \infty}(s\bar{\theta})$ gives us the imposed initial field (=0), note that:

$$\lim_{s \to 0} \frac{(1 - e^{-m_0 z})}{s} = \frac{z}{Pe}; \quad \lim_{s \to \infty} \frac{(1 - e^{-m_0 z})}{s} = 0 \tag{A-18}$$

Notice also that the series in (A-17) is quickly converging. Normally summing very few terms would be enough. To get the compact model only 1 term will be retained.

## Appendix B  Coefficients of the ODE

$$A_n = n^2\pi^2 + s \tag{B-1}$$

$$S_n = \left[\frac{2}{(1+\delta_{n0})}\right] \int_0^1 \cos(n\pi y) \left[s\bar{T}_{QS} - (T_0 - T_{QS}|_{t=0})\right] dy \tag{B-2}$$

$$B_{mn} = \left[\frac{2}{(1+\delta_{n0})}\right] \int_0^1 \cos(m\pi y)\, w(y)\, \cos(n\pi y)\, dy \tag{B-3}$$

Matrix $B_{mn}$ is composed of numbers only and is given below (Table 1) for the first 5 rows and columns. It is a symmetric matrix, except for the first row, (outside the diagonal element) which is double the first column. Diagonal elements are almost unity, while off-diagonal elements are smaller, and decrease as we go away from the diagonal. It is worthwhile noting that if the velocity profile was flat, $w=1$, then $B_{mn}$ would be the identity matrix. The parabolic velocity profile modifies the matrix a bit, to introduce coupling between different modes, through off-diagonal terms, although the coupling is rather weak.

*Table 1 Elements of the matrix $B_{mn}$ for a parabolic velocity profile*

| m \ n | 0 | 1 | 2 | 3 | 4 | 5 |
|---|---|---|---|---|---|---|
| 0 | 1.00000 | 0.60793 | -0.15198 | 0.06755 | -0.03800 | 0.02432 |
| 1 | 0.30396 | 0.92401 | 0.33774 | -0.09499 | 0.04593 | -0.02744 |
| 2 | -0.07599 | 0.33774 | 0.98100 | 0.31612 | -0.08443 | 0.03998 |
| 3 | 0.03377 | -0.09499 | 0.31612 | 0.99156 | 0.31017 | -0.08074 |
| 4 | -0.01900 | 0.04593 | -0.08443 | 0.31017 | 0.99525 | 0.30772 |
| 5 | 0.01216 | -0.02744 | 0.03998 | -0.08074 | 0.30772 | 0.99696 |

As for $S_n$, they are reported below (Table 2) for both parabolic and flat profiles, assuming zero initial condition ($T_0 - T_{QS}|_{t=0} = 0$):

*Table 2- Values of $S_n$ for parabolic and flat velocity profiles*

| N | 0 | 1 | 2 | 3 | 4 |
|---|---|---|---|---|---|
| $S_n$ parabolic | $0.085714 + q_{in,b} + z/Pe$ | -0.26424 | 0.05451 | -0.02328 | 0.012906 |
| $S_n$ flat | $q_{in,b} + z/Pe$ | -0.20264 | 0.05066 | -0.02252 | 0.012665 |

For both cases, parabolic and flat velocity profiles, the sequence of $S_n$ terms is dominated by the zeroth and first order terms. The remaining terms are not only small but are much closer to each other for the two cases.

## Appendix C  Approximation of the exponent

Let us be inspired by an approximation to $m_0$ (the case n=0 is the most important term in the expansion) and extend it to $m_n$ where $n>0$. In fact,

$$m_0 z = \left[\left(\left(\frac{Pe}{2}\right)^2 + s\right)^{\frac{1}{2}} - \left(\frac{Pe}{2}\right)\right] z \approx \left(\left(\frac{s}{Pe^2}\right)\right)(zPe),$$ and by extension we can write:

$$m_n z = \left[\left(\left(\frac{Pe}{2}\right)^2 + n^2 \pi^2 + s\right)^{\frac{1}{2}} - \left(\frac{Pe}{2}\right)\right] z \approx \left(\left(\frac{n^2 \pi^2 + s}{Pe^2}\right)\right)(z\,Pe).$$

The well-known first order Pade approximation for small $x$ is:

$$e^{-x} \approx \frac{1}{1+x}$$